# Characterization of dielectric barrier discharge in air applying current measurement, numerical simulation and emission spectroscopy

**Priyadarshini Rajasekaran, Nikita Bibinov and Peter Awakowicz**

Institute for Electrical Engineering and Plasma Technology, Ruhr-Universität Bochum, Universitätstr. 150, 44801 Bochum, Germany

E-mail: rajasekaran@aept.rub.de, nikita.bibinov@rub.de, awakowicz@aept.rub.de

**Abstract.** Dielectric barrier discharge (DBD) in air is characterized applying current measurement, numerical simulation and optical emission spectroscopy (OES). For OES, a non-calibrated spectrometer is used. This diagnostic method is applicable when cross-sectional area of the active plasma volume and current density can be determined. The nitrogen emission in the spectral range of 380 nm- 406 nm is used for OES diagnostics. Electric field in the active plasma volume is determined applying the measured spectrum, well-known Frank-Condon factors for nitrogen transitions and numerically- simulated electron distribution functions. The measured electric current density is used for determination of electron density in plasma. Using the determined plasma parameters, the dissociation rate of nitrogen and oxygen in active plasma volume are calculated, which can be used by simulation of the chemical kinetics.

## 1. Introduction

Dielectric barrier discharge (DBD) in air has a wide field of application because of the simplicity of electrical scheme of excitation, consumption of ambient air at atmospheric pressure as working gas and low averaged electric current. This discharge can, for example, activate surfaces of different materials before painting or produce biologically-active molecules like nitric oxide and ozone close to the treated surface (such as human skin) that are useful for skin therapy.

For safe and effective application of the DBD, optimisation of treatment process requires discharge characterisation or in other words, the determination of the gas temperature and plasma parameters (electron distribution function EDF and electron density). Applying these data, the plasma chemical kinetics can be simulated and the treatment conditions can be optimized. Several experimental methods such optical emission spectroscopy (OES), measurement of V-I characteristics and microphotography, and simulations are necessary for the characterization of plasma devices such as the DBD [1].

OES in combination with numerical simulation is used for the determination of gas temperature, electron velocity distribution function ($f(E)$ in eV$^{-3/2}$) and electron density ($n_e$ in m$^{-3}$) of different discharge modes of a DBD in air [1,2,3]. The electric field is determined from the intensity ratio of spectral bands of nitrogen [4]. For the determination of $n_e$, the absolute intensity of nitrogen emission is used, and for this absolutely-calibrated spectrometers are employed. The calibration of spectral devices is not always feasible due to the requirement of related equipments (such as tungsten-ribbon lamp [5]), and lack of expertise.

In this paper, we present a method of characterizing the DBD in air by implementing a *non-calibrated* spectrometer for measuring the emission spectra. In addition, current measurements and numerical simulation are used. Electric current is a function of both the electric field and the electron density, and therefore current measurements can deliver additional information that complement OES [6]. The electric field in the active plasma is determined from the intensity ratio of $N_2^+$(B-X,0-0) and $N_2$(C-B,0-0) while the electron density at that electric field is determined from the measured electric current density.

## 2. DBD device and plasma diagnostics

*2.1 DBD device*

The DBD device used in this work [1] consists of a ring-shaped copper electrode (diameter = 8 mm) covered with ceramic ($Al_2O_3$) of 1 mm thickness. Objects of high capacitance or grounded electrodes of different materials and profiles (flat or pointed) can serve as the counter electrode. A high voltage pulsed supply (trigger frequency = 300 Hz; maximum amplitude =12 kV [1]) is applied on the DBD electrode to ignite plasma in ambient air between the electrodes. Each trigger pulse initiates a sequence of high voltage pulses and damped oscillations; the frequency within each sequence is 100 kHz. At the same experimental conditions such as applied voltage, trigger frequency and inter-electrode distance, a homogeneous discharge and a stochastic-filamentary discharge are obtained, respectively, with glass and aluminium as the grounded electrode [3]. A single-filamentary discharge [2] is ignited when a pointed electrode such as a grounded spike is used. In this work, the homogeneous DBD is considered for discussion. The gap between the DBD-electrode and glass plate is 3.0 mm.

*2.2 Plasma diagnostics*

Optical emission spectroscopy is performed using a grating spectrometer (Ocean Optics QE65000, spectral range = 330-410 nm, spectral resolution ~ 0.15 nm) calibrated in wavelength by the manufacturer but the spectral efficiency remains unknown. The observed spectrum of plasma in air ($I_\lambda$ in counts·nm$^{-1}$·s$^{-1}$) shows bands of neutral and ionic emissions of molecular nitrogen (figure 1). The discharge current is measured using a current monitor (Pearson 6223). The traces are recorded using an oscilloscope (LeCroy Waverunner 204 Xi - A, 2 GHz) (figure 2). The averaged plasma duration is determined at FWHM (full width at half maximum) of the measured current profile which amounts to 20 ns, and the current-amplitude is 0.94 A.

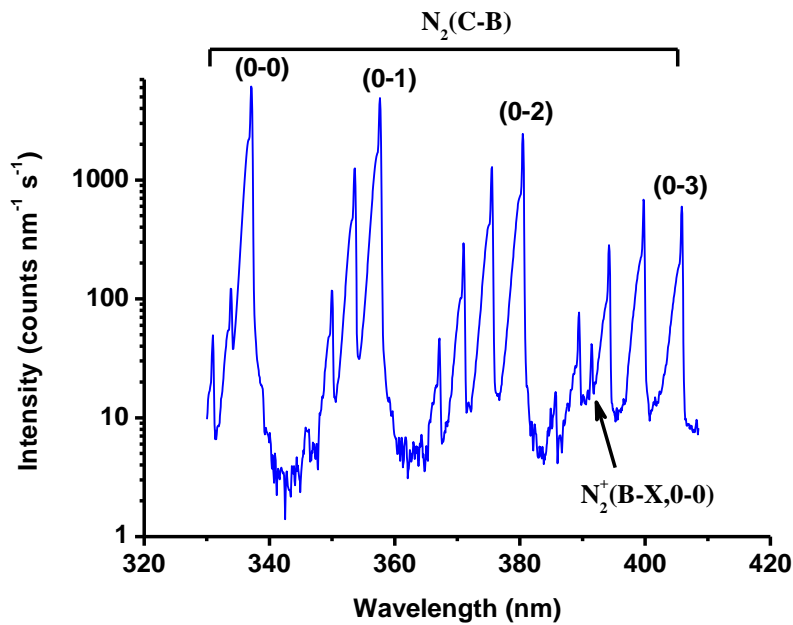

**Figure 1.** Emission spectrum of air-DBD showing different vibrational bands of neutral nitrogen molecules $N_2$(C-B), and $N_2^+$(B-X,0-0) at 391 nm.

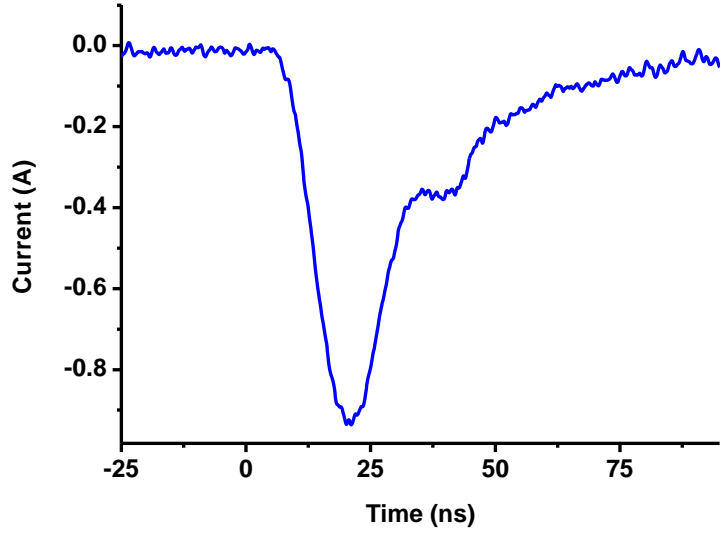

**Figure 2.** Measured current in homogeneous DBD in air with glass as the counter electrode. Inter-electrode distance amounts to 3.0 mm.

**3. Plasma characterization**

Plasma characterization refers to the determination of gas temperature and plasma parameters namely the electron distribution function and electron density. The gas temperature ($T_g$ in K) in the active plasma channel is determined using emission of $N_2$(C-B,0-0) at 337.1 nm with the assumption that the rotational temperature ($T_{rot}$) of diatomic molecules is equal to the gas temperature at atmospheric-pressure condition. Electron distribution function is simulated using reduced electric field- determined using measured intensity ratio of $N_2$(C-B) and $N_2^+$(B-X) emissions. Electron density is determined using the measured electric current density and the drift velocity calculated for the determined electric field. The gas temperature and the plasma parameters are all averaged over time and space.

*3.1. Determination of electric field from intensity ratio*

In our previous work [6], it has been shown that only direct excitation of $N_2$(C-B,0-0) at 337.1 nm and $N_2^+$(B-X,0-0) at 391.4 nm due to electron-impact of ground state neutrals $N_2$(X) occur in the homogeneous mode of the DBD in air [3], and the step wise excitation via nitrogen metastables $N_2$(A) and ground state of molecular ion $N_2^+$(X) can be neglected. Accordingly, the intensity ratio of these emissions is written as:

$$\frac{I_{N_2^+(B-X,0-0)}}{I_{N_2(C-B,0-0)}} = \frac{Q_{N_2^+(B)} \cdot N_{N_2} \cdot k_{N_2^+(B)} \cdot n_e}{Q_{N_2(C)} \cdot N_{N_2} \cdot k_{N_2(C)} \cdot n_e} = \frac{Q_{N_2^+(B)} \cdot k_{N_2^+(B)}}{Q_{N_2(C)} \cdot k_{N_2(C)}} \tag{1}$$

where, $Q_{N_2^+(B)} = \dfrac{A'}{A' + k_{qN_2}^{N_2^+(B)} \cdot N_{N_2} + k_{qO_2}^{N_2^+(B)} \cdot N_{O_2}}$ and $Q_{N_2(C)} = \dfrac{A}{A + k_{qN_2}^{N_2(C)} \cdot N_{N_2} + k_{qO_2}^{N_2(C)} \cdot N_{O_2}}$,

A' and A are the corresponding Einstein coefficients [7], $k_{qN_2}^{N_2^+(B)}$, $k_{qN_2}^{N_2(C)}$, $k_{qO_2}^{N_2^+(B)}$, $k_{qO_2}^{N_2(C)}$ (in $m^3 \cdot s^{-1}$) are the rate constants for quenching of $N_2^+(B)$ and $N_2(C)$ excited states during collision with $N_2$ and $O_2$ [5], $N_{N_2}$ and $N_{O_2}$ (in $m^{-3}$) are the densities of $N_2$ and $O_2$ at determined gas temperature, $k_{N_2^+(B)}$ and $k_{N_2(C)}$ (in $m^3 \cdot s^{-1}$) are the rate constants, respectively, for electron impact excitation of $N_2$(C-B,0-0) and $N_2^+$(B-X,0-0) emission from $N_2(X)$, and $n_e$ (in $m^{-3}$) is the electron density. The quenching factors $Q_{N_2^+(B)}$ and $Q_{N_2(C)}$ depend slightly on the gas temperature because of temperature dependencies of gas density and quenching rate constants according to Arrhenius formula. This effect will be discussed later.

The rate constants for electron-impact excitation of nitrogen emissions depend on the electron velocity distribution function ($f_v(E)$ in $eV^{-3/2}$) and the corresponding cross-section σ (in $m^2$) [8]. $k_{N_2^+(B)}$ and $k_{N_2(C)}$ are calculated using (2) for varied electric field values:

$$k_{exc} = 4\pi\sqrt{2} \int_0^\infty f_v(E) \sqrt{\dfrac{2e}{m}} E \cdot \sigma_{exc}(E) \, dE, \qquad (2)$$

where, e and m are elementary charge (in C) and mass of an electron (in kg). Here, E is the kinetic energy of electrons (in eV). $f_v(E)$ is normalized to fulfil (3):

$$4\pi\sqrt{2} \int_0^\infty f_v(E) \sqrt{E} \, dE = 1 \qquad (3)$$

$f_v(E)$ is simulated by solving the Boltzmann equation in 'local approximation' (nitrogen/ oxygen = 78%/ 21%) for varied electric field values [9]. The program code 'EEDF' developed by Prof. A P Napartovich [10] is used for this purpose. $k_{N_2^+(B)}$, $k_{N_2(C)}$ and the ratio between these rate constants for discharge in air are listed in table 1.

| Reduced electric field E/N in Td | Electron-impact excitation rate constant for | | Ratio of excitation rates of nitrogen emissions excluding quenching |
|---|---|---|---|
| | $N_2^+$(B-X,0-0) $k_{exc}^{N_2^+(B)}$ in $m^3 \cdot s^{-1}$ | $N_2$(C-B,0-0) $k_{exc}^{N_2(C)}$ in $m^3 \cdot s^{-1}$ | $\dfrac{k_{exc}^{N_2^+(B)}}{k_{exc}^{N_2(C)}}$ |
| 20 | 1.55E-35 | 2.70E-24 | 5.74E-12 |
| 40 | 6.56E-27 | 1.74E-20 | 3.77E-07 |
| 60 | 1.51E-23 | 6.06E-19 | 2.49E-05 |
| 80 | 9.10E-22 | 3.95E-18 | 2.30E-04 |
| 100 | 1.13E-20 | 1.22E-17 | 9.26E-04 |
| 120 | 6.30E-20 | 2.58E-17 | 2.44E-03 |
| 140 | 2.21E-19 | 4.40E-17 | 5.02E-03 |
| 160 | 5.80E-19 | 6.57E-17 | 8.83E-03 |
| 180 | 1.25E-18 | 8.97E-17 | 1.39E-02 |
| 200 | 2.37E-18 | 1.15E-16 | 2.06E-02 |
| 220 | 4.05E-18 | 1.41E-16 | 2.87E-02 |
| 240 | 6.41E-18 | 1.68E-16 | 3.82E-02 |
| 260 | 9.57E-18 | 1.95E-16 | 4.91E-02 |
| 280 | 1.36E-17 | 2.21E-16 | 6.15E-02 |
| 300 | 1.86E-17 | 2.46E-16 | 7.56E-02 |
| 320 | 2.46E-17 | 2.71E-16 | 9.08E-02 |
| 340 | 3.17E-17 | 2.95E-16 | 1.07E-01 |
| 360 | 3.98E-17 | 3.17E-16 | 1.26E-01 |
| 380 | 4.91E-17 | 3.39E-16 | 1.45E-01 |
| 400 | 5.94E-17 | 3.60E-16 | 1.65E-01 |
| 420 | 7.07E-17 | 3.79E-16 | 1.87E-01 |
| 440 | 8.30E-17 | 3.97E-16 | 2.09E-01 |
| 460 | 9.62E-17 | 4.15E-16 | 2.32E-01 |
| 480 | 1.10E-16 | 4.31E-16 | 2.55E-01 |
| 500 | 1.24E-16 | 4.46E-16 | 2.78E-01 |

**Table 1.** Ratio of nitrogen emissions (excluding quenching) determined using (3) for different electric field values at atmospheric pressure in air.

When using a calibrated spectrometer, the intensities of nitrogen emission $N_2$(C-B,0-0) and $N_2^+$(B-X,0-0) are calculated by integration of measured intensities ($I_\lambda$) corrected to efficiency of spectrometer ($\varepsilon_\lambda^{-1}$ in photons per count) (4). The measured spectrum is integrated in $\lambda_1$ to $\lambda_2$ and $\lambda_3$ to $\lambda_4$ spectral ranges, respectively, for determination of $N_2^+$(B-X,0-0) (390..392 nm) and $N_2$(C-B,0-0) (333...337.5 nm) intensities. The efficiency of the spectrometer is a smooth function of wavelength and can be presented as constants ($\varepsilon_{391}^{-1}, \varepsilon_{337}^{-1}$) in short integrating intervals.

$$\frac{I_{N_2^+(B-X,0-0)}}{I_{N_2(C-B,0-0)}} = \frac{\int_{\lambda_1}^{\lambda_2} I_\lambda \cdot \varepsilon_\lambda^{-1} d\lambda}{\int_{\lambda_3}^{\lambda_4} I_\lambda \cdot \varepsilon_\lambda^{-1} d\lambda} = \frac{\varepsilon_{391}^{-1} \cdot \int_{\lambda_1}^{\lambda_2} I_\lambda d\lambda}{\varepsilon_{337}^{-1} \cdot \int_{\lambda_3}^{\lambda_4} I_\lambda d\lambda} \qquad (4)$$

But the spectrometer used in this experiment is not calibrated, and therefore for the use of equation (4) one needs additional information and assumptions. We assume that the efficiency is a linear function of wavelength and we find the slope of this function. But to determine ratio of efficiency values at $\lambda$ = 337 nm and $\lambda$ = 391 nm, linearity in a such a broad spectral range cannot be valid for the spectrometer used. To solve this problem, we use emission bands of $N_2$(C-B,0-2) at $\lambda$ = 380 nm and $N_2$(C-B,0-3) at $\lambda$ = 406 nm that are close to $N_2^+$(B-X,0-0) at $\lambda$ = 391 nm and linearity of efficiency curve of spectrometer in the range of 380-406 nm is valid. We integrate emission spectrum in spectral rages $\lambda_5$ to $\lambda_6$ (376.5 to 381.5 nm and $\lambda_7$ to $\lambda_8$), and $\lambda_7$ to $\lambda_8$ (401 to 406.5 nm) correspondingly for $N_2$(C-B,0-2) and $N_2$(C-B,0-3). In order to determine the intensity of $N_2$(C-B,0-0), the calculated intensities of $N_2$(C-B,0-2) and $N_2$(C-B,0-3) are corrected to the ratio of corresponding Frank-Condon factors (5). The branching factors of emission from excited vibrational level $N_2$(C,0) that corresponds to the Frank-Condon factors are well known [11] and independent of plasma conditions.

$$\frac{I_{N_2^+(B-X,0-0)}}{I_{N_2(C-B,0-0)}} = \frac{\varepsilon_{391}^{-1} \cdot \int_{\lambda_1}^{\lambda_2} I_\lambda d\lambda}{\varepsilon_{380}^{-1} \cdot \frac{FCF_{0-0}}{FCF_{0-2}} \cdot \int_{\lambda_5}^{\lambda_6} I_\lambda d\lambda} = \frac{\varepsilon_{391}^{-1} \cdot \int_{\lambda_1}^{\lambda_2} I_\lambda d\lambda}{\varepsilon_{406}^{-1} \cdot \frac{FCF_{0-0}}{FCF_{0-3}} \cdot \int_{\lambda_7}^{\lambda_8} I_\lambda d\lambda} = \frac{\varepsilon_{391}^{-1} \cdot R_{380}}{\varepsilon_{380}^{-1}} = \frac{\varepsilon_{391}^{-1} \cdot R_{406}}{\varepsilon_{406}^{-1}}, \qquad (5)$$

where,

$$R_{380} = \frac{\int_{\lambda_1}^{\lambda_2} I_\lambda d\lambda}{\frac{FCF_{0-0}}{FCF_{0-2}} \cdot \int_{\lambda_5}^{\lambda_6} I_\lambda d\lambda} = 0.272 \cdot \frac{\int_{\lambda_1}^{\lambda_2} I_\lambda d\lambda}{\int_{\lambda_5}^{\lambda_6} I_\lambda d\lambda} \quad \text{and} \quad R_{406} = \frac{\int_{\lambda_1}^{\lambda_2} I_\lambda d\lambda}{\frac{FCF_{0-0}}{FCF_{0-3}} \cdot \int_{\lambda_7}^{\lambda_8} I_\lambda d\lambda} = 0.084 \cdot \frac{\int_{\lambda_1}^{\lambda_2} I_\lambda d\lambda}{\int_{\lambda_7}^{\lambda_8} I_\lambda d\lambda}$$

In our opinion, the efficiency of usual UV/VIS gratings spectrometer can be, with good accuracy, assumed as linear (6) in the spectral range of 380-406 nm. Taking into account (5), we receive (7).

$$\varepsilon_{391}^{-1} = \varepsilon_{380}^{-1} + 0.42 \cdot (\varepsilon_{406}^{-1} - \varepsilon_{380}^{-1}) = \varepsilon_{380}^{-1}\left(1 + 0.42 \cdot \left(\frac{\varepsilon_{406}^{-1}}{\varepsilon_{380}^{-1}} - 1\right)\right) = \varepsilon_{380}^{-1}\left(1 + 0.42 \cdot \left(\frac{R_{406}}{R_{380}} - 1\right)\right) \quad (6)$$

$$\frac{I_{N_2^+(B-X,0-0)}}{I_{N_2(C-B,0-0)}} = \frac{Q_{N_2^+(B)} \cdot k_{N_2^+(B)}}{Q_{N_2(C)} \cdot k_{N_2(C)}} = R_{380} + 0.42 \cdot (R_{406} - R_{380}) \quad (7)$$

Applying (7) and (4), we determine electron distribution function using the non-calibrated spectrometer.

*3.2. Measured current as a diagnostic tool*

In the DBD, electric current (*i*) flows between the electrodes through a defined plasma volume with cross section S and there is no current loss to the surrounding. This allows the use of measured current as a diagnostic tool for plasma characterization. The current density (j in A·m$^{-2}$) is a function of electron density and drift velocity ($v_d$ in m·s$^{-1}$) of electrons (8), which in turn is a function of electric field.

$$j = \frac{i}{S} = n_e \cdot e \cdot v_d \quad (8)$$

$v_d$ determined for different electric field values and measured current density are used in (9) to determine $n_e$ which itself becomes a function of the electric field:

$$n_e = = \frac{j}{e \cdot v_d} = F(E/N) \quad (9)$$

The electron drift velocity used in (9) to determine $n_e$ for different electric field values is presented in table 2. As was shown [6], a combination of OES and current measurement allows determination of electron distribution function and electron density in the active plasma volume.

**Table 2**. Electron drift velocity ($v_d$ in m·s$^{-1}$) for different electric field values used for the determination of electron density.

| Reduced electric field E/N in Td | Electron drift velocity $v_d$ in m·s$^{-1}$ |
|---|---|
| 20 | 3.34E4 |
| 40 | 5.65E4 |
| 60 | 7.48E4 |
| 80 | 9.28E4 |
| 100 | 1.11E5 |
| 120 | 1.29E5 |
| 140 | 1.47E5 |
| 160 | 1.64E5 |
| 180 | 1.80E5 |
| 200 | 1.96E5 |
| 220 | 2.12E5 |
| 240 | 2.27E5 |
| 260 | 2.42E5 |
| 280 | 2.57E5 |
| 300 | 2.71E5 |
| 320 | 2.85E5 |
| 340 | 2.98E5 |
| 360 | 3.12E5 |
| 380 | 3.24E5 |
| 400 | 3.37E5 |
| 420 | 3.49E5 |
| 440 | 3.61E5 |
| 460 | 3.72E5 |
| 480 | 3.83E5 |
| 500 | 3.94E5 |

## 4. Results and Discussion

To determine gas density in the plasma and the rate constants of reactions between gas species, we determine the gas temperature at DBD conditions. At that rotational distribution in $N_2$(C-B,0-0) emission (figure 3) is applied. The emission spectrum of $N_2$(C-B,0-0) is simulated at variable rotational temperature and compared with the measured one. Rotational temperature of nitrogen molecules, which is equal to the gas temperature at atmospheric pressure conditions, is determined in fitting procedure. Applying this method, $T_g$ in the homogeneous DBD is determined as $360 \pm 30$ K.

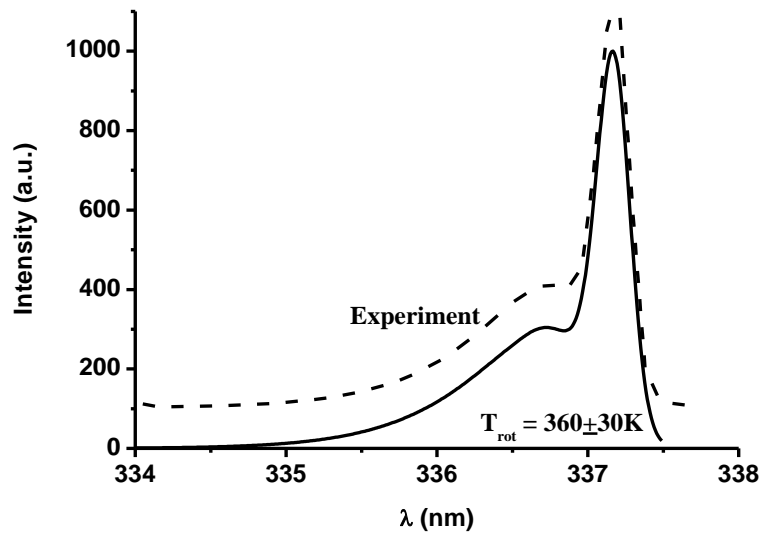

**Figure 3.** $N_2$(C-B,0-0) emission band measured (broken line) and simulated (solid line) at 360K. The measured spectrum is shifted for clarity.

Using (4), we calculate the ratio of intensities of nitrogen emissions at 360 K and DBD conditions in air (table 3). These calculated values are also presented in figure 4. Using this data and measured emissions, the ratio of nitrogen emissions calculated using (8) amounts to $4.54 \cdot 10^{-3}$ that corresponds to 224 Td which is the electric field at the studied DBD condition.

**Table 3.** Ratio of observed nitrogen emissions determined using (4) for different electric field values in air at atmospheric pressure

| Reduced electric field E/N in Td | Ratio of observed nitrogen emissions in air at 360 K in atmospheric pressure $\frac{I_{N_2^+(B-X,0-0)}}{I_{N_2(C-B,0-0)}} = \frac{k_{exc}^{N_2^+(B)} \cdot Q_{N_2^+(B)}}{k_{exc}^{N_2(C)} \cdot Q_{N_2(C)}}$ |
|---|---|
| 20 | 8.76E-13 |
| 40 | 5.76E-8 |
| 60 | 3.80E-6 |
| 80 | 3.51E-5 |
| 100 | 1.41E-4 |
| 120 | 3.73E-4 |
| 140 | 7.66E-4 |
| 160 | 1.35E-3 |
| 180 | 2.12E-3 |
| 200 | 3.15E-3 |
| 220 | 4.38E-3 |
| 240 | 5.83E-3 |
| 260 | 7.50E-3 |
| 280 | 9.39E-3 |
| 300 | 1.15E-2 |
| 320 | 1.39E-2 |
| 340 | 1.63E-2 |
| 360 | 1.92E-2 |
| 380 | 2.21E-2 |
| 400 | 2.52E-2 |
| 420 | 2.86E-2 |
| 440 | 3.19E-2 |
| 460 | 3.54E-2 |
| 480 | 3.89E-2 |
| 500 | 4.24E-2 |

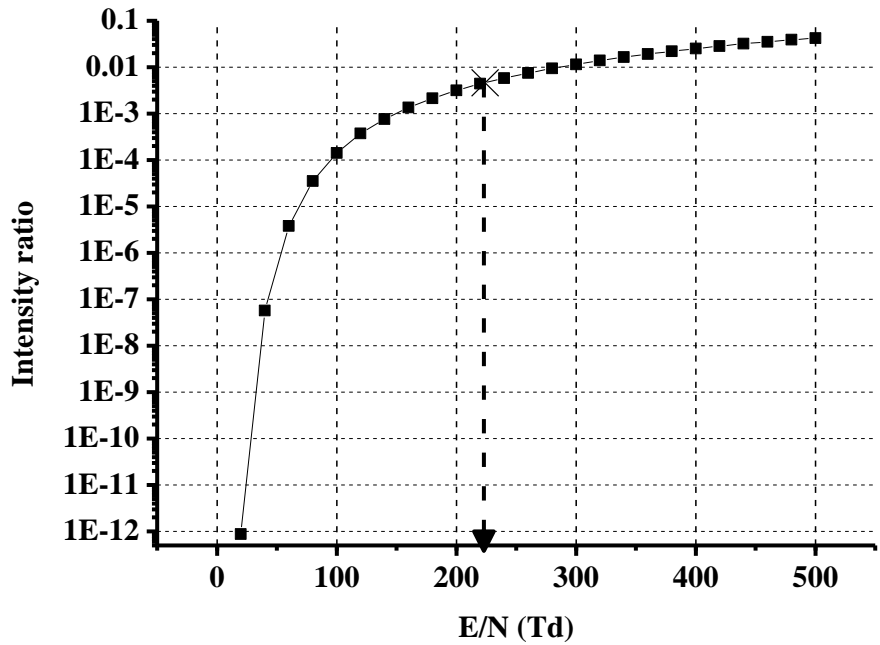

**Figure 4.** Ratio of nitrogen emissions (including quenching at $T_g=360K$ determined using (4) for different electric field values. The data pertaining to this curve is also presented in table 3. The point (×) corresponds to measured ratio of intensities of nitrogen emissions in the homogeneous DBD reported here.

At our experimental conditions, the homogeneous DBD fills the entire gap between electrodes, and hence the circular cross-section of the active plasma volume amounts to $7.85 \cdot 10^{-5}$ m$^2$. The averaged electric current is 0.47 A and the current density amounts to $5.9 \cdot 10^3$ A·m$^{-2}$. Applying graphical method for (4) and (9) (figure 5), we determine the averaged electron density as $3.6 \cdot 10^{17}$ m$^{-3}$ at 224 Td in the homogeneous mode. These values are in good agreement with those determined using absolutely and relatively-calibrated spectrometer [6].

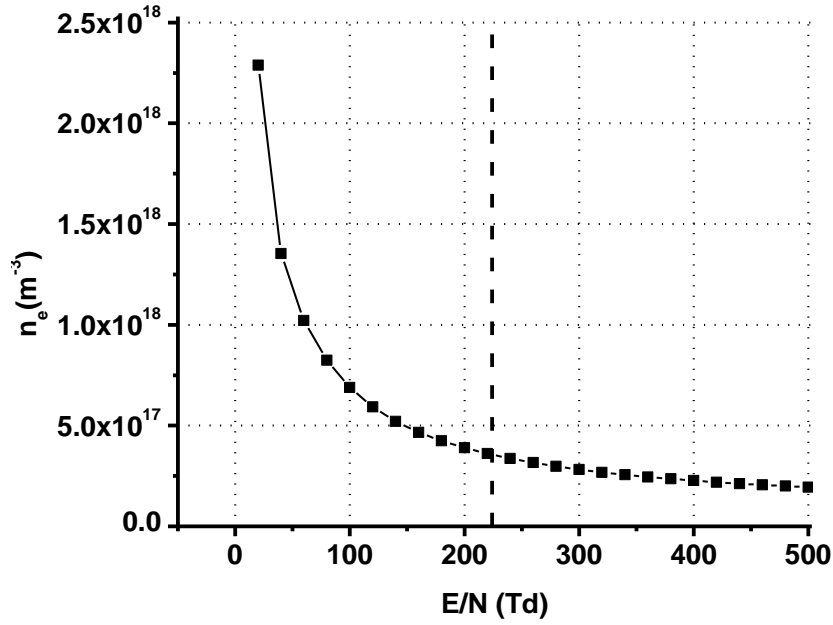

**Figure 5**. Graphical interpretation of equations (7,9) for homogenous DBD. Broken line presents electric field (E/N = 224 Td) determined applying ratio of nitrogen intensities (7). ■ - applying measured current density (9). $n_e^{224Td} = 3.6 \cdot 10^{17}$ m$^{-3}$.

To determine the influence of gas temperature on the calculated ratio of the quenching factors, we calculate the ratio $\dfrac{Q_{N_2^+(B)}}{Q_{N_2(C)}}$ at variable gas temperatures. This ratio is equal to 6.55 at 300 K and decreases to 0.7 % (of 6.55) at 1000 K. Therefore, we conclude that the influence of gas temperature on plasma parameters is negligible. However, the influence of gas temperature cannot be neglected while determining the rate of production of nitrogen and oxygen atoms during electron-impact dissociation of their molecules because the density of gas species ($N_{N_2}$ and $N_{O_2}$) depend on the gas temperature. Furthermore, the rate constants of chemical reactions also depend on the gas temperature. Therefore, for the simulation of plasma chemical kinetics, the determination of gas temperature is essential.

The plasma parameters and the gas temperature can be used for the simulation of chemical kinetics and the determination of species' fluxes reaching the treated surface. The rate constants for electron-impact dissociation of nitrogen and oxygen molecules in air ($k_{diss}^{N_2}$ and $k_{diss}^{O_2}$ in m$^3$·s$^{-1}$) for varied electric fields are presented in table 4. The dissociation rates ($R_{diss.}^{M}$ in m$^{-3}$·s$^{-1}$) of nitrogen and oxygen molecules at plasma conditions are calculated using (10):

$$R_{diss}^{M} = N_M \cdot k_{diss}^{M} \cdot n_e \qquad (10)$$

where, M = N$_2$ or O$_2$ correspondingly. In the homogeneous discharge at DBD conditions (E/N=224 Td, n$_e$=3.6·10$^{17}$ m$^{-3}$), $R_{diss.}^{N_2}$ = 7.98·10$^{27}$ m$^{-3}$·s$^{-1}$ and $R_{diss.}^{O_2}$ = 1.12·10$^{27}$ m$^{-3}$·s$^{-1}$.

**Table 4.** Rate constants for electron impact dissociation of nitrogen and oxygen by variation of electric field.

| Reduced electric field E/N in Td | Dissociation rate constants by electron impact | |
|---|---|---|
| | N$_2$+e→2N +e $k_{diss}^{N_2}$ in m$^3$·s$^{-1}$ | O$_2$+e→2O +e $k_{diss}^{O_2}$ in m$^3$·s$^{-1}$ |
| 20 | 4.54E-22 | 2.87E-20 |
| 40 | 5.39E-19 | 3.73E-18 |
| 60 | 9.97E-18 | 3.02E-17 |
| 80 | 4.79E-17 | 9.20E-17 |
| 100 | 1.26E-16 | 1.77E-16 |
| 120 | 2.46E-16 | 2.72E-16 |
| 140 | 4.05E-16 | 3.67E-16 |
| 160 | 5.99E-16 | 4.58E-16 |
| 180 | 8.25E-16 | 5.45E-16 |
| 200 | 1.08E-15 | 6.27E-16 |
| 220 | 1.36E-15 | 7.03E-16 |
| 240 | 1.67E-15 | 7.76E-16 |
| 260 | 2.00E-15 | 8.43E-16 |
| 280 | 2.35E-15 | 9.05E-16 |
| 300 | 2.73E-15 | 9.64E-16 |
| 320 | 3.11E-15 | 1.01E-15 |
| 340 | 3.52E-15 | 1.06E-15 |
| 360 | 3.93E-15 | 1.11E-15 |
| 380 | 4.36E-15 | 1.16E-15 |
| 400 | 4.80E-15 | 1.20E-15 |
| 420 | 5.24E-15 | 1.23E-15 |
| 440 | 5.69E-15 | 1.27E-15 |
| 460 | 6.14E-15 | 1.30E-15 |
| 480 | 6.59E-15 | 1.33E-15 |
| 500 | 7.04E-15 | 1.36E-15 |

The method of using a non-calibrated spectrometer for characterization of discharges in air at atmospheric-pressure conditions is applied in assumption of linearity of the efficiency function of the spectrometer in spectral range of about 25 nm. The method discussed in this work can be applied when a usual grating spectrometer is used. By application of an echelle spectrometer operated in numerous optical orders that possesses efficiency curve which is "rugged" profile with sharp contour that cannot be described as a linear function of wavelength also by such small spectral range, this method is not applicable.

To numerically-simulate emission spectrum at different gas temperature that can be used for the determination of gas temperature in the experiment, the program code used was developed by our group specially for this purpose [12]. Similar simulation can also be performed using applications like "Specair" and "LIFBASE". To simulate electron distribution function and to calculate the drift velocity of electrons, the Boltzmann equation can be numerically solved using the program "BOLSIG". (These applications and programs are available as freeware that can be downloaded from their respective websites)

**5. Conclusion**

The combination of current measurement, numerical simulation and OES has been applied for the determination of plasma parameters at DBD conditions in air. For OES, a non-calibrated spectrometer is used. Electric field and electron distribution function are determined using the measured unaltered emission spectrum of nitrogen, the well-known Frank-Condon factors of nitrogen emission and the calculated rate constants for electron-impact excitation of these emissions. Electron density is determined using measured electric current density and drift velocity that depend on the electric field. The determined plasma parameters can be used for calculation of dissociation rates at DBD conditions and for simulation of plasma chemical kinetics.